The potential of mass rearing of ***Monoska dorsiplana*** (Pteromalidae) a native gregarious ectoparasitoid of ***Pseudopachymeria spinipes*** (Bruchidae) in South America.

**Danielle Rojas-Rousse[1], Karine Poitrineau and César Basso ***

Institut de Recherche sur la Biologie des Insectes (IRBI), UMR du CNRS 6035

Faculté des Sciences et Techniques, Avenue Monge, 37200 – Tours – France

* Facultad de Agronomía. Av. Garzón 780. 12900 Montevideo. Uruguay.

1-Corresponding author : rousse@univ-tours.fr

Phone number: 02 47 36 69 73

FAX number: 02 47 36 69 66

**Abstract**

In Chile and Uruguay, the gregarious Pteromalidae (*Monoska dorsiplana)* has been discovered emerging from seeds of the persistent pods of *Acacia caven* attacked by the univoltin bruchid *Pseudopachymeria spinipes*. We investigated the potential for mass rearing of this gregarious ectoparasitoid on an alternative bruchid host, *Callosobruchus maculatus,* to use it against the bruchidae of native and cultured species of Leguminosea seeds in South America.

The mass rearing of *M. dorsiplana* was carried out in a population cage where the density of egg-laying females per infested seed was increased from 1:1 on the first day to 5:1 on the last (fifth) day. Under these experimental conditions egg-clutch size per host increased, and at the same time the mortality of eggs laid also increased. The density of egg-laying females influenced the sex ratio which tended towards a balance of sons and daughters, in contrast to the sex ratio of a single egg-laying female per host (1 son to 7 daughters). The mean weight of adults emerging from a parasitized host was negatively correlated with the egg-clutch size, i.e. as egg-clutch size increased, adult weight decreased.

All these results show that mass rearing of the gregarious ectoparasitoid *M. dorsiplana* was possible under laboratory conditions on an alternative bruchid host *C. maculatus*. As *M. dorsiplana* is a natural enemy of larval and pupal stages of bruchidae, the next step was to investigate whether the biological control of bruchid *C. maculatus* was possible in an experimental structure of stored beans.

**Key words. Gregarious parasitoid, egg-clutch size, theoretical offspring, observed offspring, sex ratio, bruchid host, *Callosobruchus maculatus***

## 1. Introduction

Bruchids constitute the largest single problem for native and cultured species of Leguminosea seeds in Latin America, attacking a number of economically important plant species. The common bean weevil *Acanthoscelides obtectus* (Say) and the Mexican bean weevil *Zabrotes subfasciatus* (Boh) are the main post-harvest pests of dry beans and currently constitute a major problem in the management of bean stocks in storage sites (Schmale *et al.,* 2001; Alvarez *et al.,* 2005 ). In the last 30 years, these two bruchid species have also been recorded on new host plant species, such as *Cajanus indicus*, *Pisum sativum, Vicia faba,* and *Vigna unguiculata* (Jarry and Bonet, 1982; Johnson,1983, 1990). This expansion of host range requires new integrated pest management strategies based on natural resources, including parasitoids. In South America, as in traditional storage systems in the African tropical belt, the parasitoid *Dinarmus basalis* (Ashm.) is currently the main candidate for the biological control of bruchids in stored beans (Schmale *et al.,* 2001; Sanon *et al.,* 1998; Dorn *et al.,* 2005).

The challenge now is to find one or more appropriate biological control agents which are native to Latin America. Two native Trichogrammatidae have recently been found as oophagous parasitoids of bruchid beetle eggs: *Uscana chiliensis* (Pintureau and Gering) on *Bruchus pisorum*, and *Uscana espinae* (Pintureau and Gering) on *Pseudopachymeria spinipes* (Er.), (Pintureau *et al.,* 1999). In addition, one Pteromalidae (*Monoska dorsiplana,* Boucek) and two Eulophidae (*Horismenus spp*.) have been found emerging from seeds of the persistent pods of *Acacia caven* (Mol.) contaminated by the univoltin bruchid *P. spinipes* (Rojas-Rousse, 2006). These persistent pods provide a natural reserve of parasitoids which are a potential resource for the biological control of Bruchidae. Previous investigations have shown that *Dinarmus vagabundus* and *Dinarmus basalis* (Pteromalidae), parasitoids of larval and pupal stages of bruchids, can be mass-reared on a substitution bruchid host, *Callosobruchus maculatus* (Rojas-Rousse *et al*., 1983; Rojas-Rousse *et al*., 1988). Some life history traits of *M. dorsiplana* have been investigated under laboratory conditions using the substitution bruchid host *Callosobruchus maculates*, and it was observed that with a low density of *M. dorsiplana* females per host, i.e. 1:1, the female laid one clutch of eggs during one oviposition, the parasitoid larvae developed gregariously, and the most common patriline was 1 male and 7 females (Rojas-Rousse, 2006).

The aim of the present study was to test how egg-clutch size changed in a population cage when the density of females per host was increased from 1:1 to 5:1 over 5 consecutives days. Under these controlled conditions, mass production of *M. dorsiplana* on the alternative host *C. maculatus* could be investigated. The egg and offspring clutch sizes were compared and the trade-off between egg and offspring clutch sizes was studied through experimental manipulation of the egg-clutch size.

## 2. Materials and Methods

### *2.1. Biological material.*

Host and parasitoid strains were mass-reared in a climatic chamber under conditions close to those of their zone of origin, with synchronous photo and thermo-periods: 30° / 20°C, 12h / 12h L:D, and 70% RH.

The bruchid host *C.maculatus* was mass reared in the laboratory on *Vigna radiata* (L.) Wilszek seeds. After egg-laying, the bruchid females were removed and the seeds stored until the larvae inside the seed reached the final larval or pupal stage.

Host size, determined by its developmental stage, is one of the main factors contributing to variations in egg-clutch size, and therefore only the largest *C. maculatus* hosts were presented to the egg-laying *M. dorsiplana* females (Terrasse *et al*, 1996; Pexton and Mayhew, 2002; Pexton and Mayhew, 2005). For this, the seeds were examined under a microscope lens and only seeds with 1 to 3 hosts, i.e. the fourth-instar larvae, prepupae and pupae, were offered to the parasitoid females. Because *C.maculatus* larvae were not directly accessible to parasitoids, the female parasitoid generally introduced her ovipositor through the hole drilled by the neonatal host larvae (van Alebeek *et al*., 1993).The parasitoid females located these holes from the egg shells remaining on the seed tegument (personal observations).

*2.2. Parasitization of the substitution bruchid host C. maculatus in a population cage.*

The experiments were conducted in a special 'altuglass' population-rearing cage (40x30x25 cm) simulating a ventilated storage structure. In this cage, 120 *V. radiata* seeds with one, two or three hosts were introduced every day with 120 newly mated *M. dorsiplana* females (mating occurred immediately after emergence of females). The bruchid-infested seeds were exposed for 24h to the parasitoids and renewed daily on 5 consecutive days, unlike parasitoid females which were not removed. In this way, theoretically the density of females per seed increased from 1:1 on the first day (120 females for 120 infested seeds) to 5:1 on the last (fifth) day (120x5 females for 120 infested seeds). The seeds removed every day were divided into two sets, one with 40 and the other with 80 seeds. All the seeds in the first set were opened to investigate the parasitism of each host, and the second set was used as a control.

*2.3. Analysis of egg-clutch size, theoretical offspring, and relative mortality*

The data recorded for each opened seed included the number and developmental stage of hosts, whether the host was parasitized or not, and if so, the egg-clutch size. Each parasitized host was incubated individually in a small plastic tube (30° / 20°C, 12h / 12h L:D, and 70% RH) to identify the developmental stage of the parasitoids, the weight of each parasitoid pupa before the moult, and the number and sex of the emerging adults.

### *2.4. Observed offspring*

All the seeds of the control set were incubated individually in a small plastic tube (30° / 20°C, 12h / 12h L:D, and 70% RH). This control set was used to determine the number and sex of parasitoid adults emerging from each parasitized seed without experimental manipulation.

### *2.5. Data analysis*

For each set of seeds, the various parameters were analysed for the 5 days of activity of the parasitoid females. Seeds with one or two parasitical hosts were analysed separately. These two sets were compared with regard to the distribution of egg-clutch sizes, the offspring observed per parasitized host, the development time of each sex and the dry weights of emerging male and female parasitoids. An ANOVA was performed (XLStats 6 for Windows) to assess the intra- and inter-variability of the sets. If the variances were statistically different, the Student-t test was performed. The Chi-square test was used to evaluate the of distribution of egg-clutch sizes between the hosts in the seeds. The influence of egg-clutch size on the parasitoid adult weight was tested by a simple linear regression (XLStats 6 for Windows).

## 3. Results

### *3.1 Parasitized* hosts

In the 200 opened seeds (40 seeds per day for 5 days), there were 323 hosts. Of these seeds, 45.5% (91/200) contained one host, 47.5% (95/200) two hosts, and 7% (14/200) three hosts. Only 67% of the hosts (216/323) were actually parasitized, i.e. contained egg clutches (Table 1). The seeds with a single parasitological host per seed were 100% attacked (Table 1). Those with two hosts were attacked less, with 59.47% of hosts parasitized (113/190), and when there were three hosts, only 28.57% of the hosts (12/42) were parasitized (Table 1).

Because 3 hosts per seed were rarely observed, our analysis was restricted to a comparison of seeds enclosing one and two hosts. The percentage of parasitized hosts was significantly greater among seeds enclosing only one host (t-test for percentage comparison $t = 6.95$; at the level of significance $\alpha = 0.05$ $t_{[.05]\,\infty} = 1.96$).

### *3.2. Distribution of egg-clutch size with one parasitized host per seed*

The distribution of egg-clutch size observed per parasitized host varied from 1 to 29 eggs with the modal class from 9 to 10 eggs (Figure 1). With one host enclosed per seed, the average clutch size was 9.37 ± 1.12 eggs, and with two hosts per seed it was 8.48 ± 0.97 (mean ± standard error of the mean). The distribution of egg-clutch sizes showed no significant difference from the normal distribution and the difference between the two means was not significantly different [Kolmogorov-Smirnov test: 1 host per seed, N (6.07; 37.35),

$D=0.176 < D_{0.05} = 0.338$; 2 hosts per seed N (5.13; 32.83), $D=0.241 < D_{0.05} = 0.338$, (Student test: t =1.07 at the level of significance $\alpha = 0.05$ $t_{[.05]\infty} = 1.96$)].

*3.3 Distribution of egg-clutch size with two hosts per seed*

With two hosts per seed, the females could parasitize only one of the two hosts (Figure 1). When both hosts were parasitized, the modal class (1-2 eggs per parasitized host) corresponded to the smallest egg clutch size (Figure 1). The modal class was larger (9-10 eggs per clutch) when one of the two hosts was parasitized (Figure 1). There was a significant difference in the mean clutch size when both hosts were parasitized, 4.17 ± 1.06 (mean ± standard error of the mean), and when one of the two hosts was parasitized: 9.37 ± 1.12 eggs (Student-t test: t =5.3 at the level of significance $\alpha = 0.05$ $t_{[.05]\infty} = 1.96$). This difference was confirmed by an irregular distribution of the observed frequencies, ranging from 1-2 to 13-14 eggs per host (Chi-square test using Yates correction: $\chi^2$ calculated = 27.25: alpha = 0.05, $\chi^2_{ddl\ 6} = 12.59$).

*3.4 Theoretical offspring and sex-ratio of observed offspring with one parasitized host per seed*

As each parasitized host was incubated individually up to the adult stage, it was possible to calculate the relative mortality: number of eggs– number of emerged adults / number of eggs. The correlation between egg-clutch size and relative mortality was strong:

R=0.99, P <0.0001, with mortality rising as egg-clutch size increased, i.e. not all the eggs of one clutch would reach adulthood.

On average, 4.12 ± 0.39 males and 3.84 ± 0.28 females emerged from one parasitized host (mean ± standard error of the mean). As the variances of emerged males and females were equal, the difference observed between their means was not statistically different [ANOVA: $F_{(0.05), 1, 427}$ calculated =1.245 with P= 0.265: $F_{\text{critical value}}$ = 3.86].

*3.5 Development time*

Observations indicated that in each clutch the male(s) emerged first while the emergence of females was spread over time.The shortest time (19 days) was for males with an average of 20.88 ± 0.15 days, and the longest (30 days) for females with an average of 21.06 ± 0.19 days (mean ± standard error of the mean). Analysis of the total development time from egg to adulthood (male or female), showed that the difference observed between the means did not significantly differ [ANOVA: $F_{(0.05), 1, 378}$ calculated =1.912 with P= 0.168: $F_{\text{critical value}}$ = 3.86].

*3.6. Dry weights of males and females in each clutch*

Dry weight distribution indicated that the lowest values (from 0.1mg to 0.9 mg) were for males and the highest (up to 1.6 mg) for females.The mean dry weight of females (0.717 ± 0.05) was double that of males 0.391 ± 0.02 (mean ± standard error of the mean).The variances of these dry weights being statistically different, the difference between the mean weights of emerged females and males was statistically different [ANOVA: $F_{(0.05), 1, 378}$

calculated =151.58 with $P = 0.0001$: $F_{critical\ value} = 3.02$, Student-t test: $t = 11.7$ at the level of significance $\alpha = 0.05$ $t_{[.05]\,\infty} = 1.96$) ]. For each sex and clutch, mean adult weight and egg-clutch size were negatively correlated (Figure 2A, B). This negative correlation indicated that the mean adult weight decreased as the egg-clutch size increased.

## 4. Discussion

In this study, *M. dorsiplana* was successfully mass-reared in a population cage. With both one and two parasitological hosts per seed but only a single host actually parasitized, the most frequent egg-clutch size was 9 to 10 eggs and the largest was 29 eggs. With a density of one to five females and one parasitological host per seed, a modal class of egg-clutch size close to that observed with one egg-laying female per host was produced (Rojas-Rousse, 2006). The smallest egg-clutch size (1 or 2 eggs) was observed when two parasitological hosts per seed were parasitized. In this situation, egg-laying was disturbed by numerous contacts between the females (personal observations).

In theory, the number of eggs laid on a host's body corresponds to the number of offspring. However, this theoretical offspring clutch size differed significantly from the actual offspring numbers emerging from parasitized hosts in the control group, indicating that not all the eggs reached the adult stage. The correlation between egg-clutch size and relative mortality was high ($R=0.99$, $P <0.0001$), with mortality rising as the egg-clutch size increased. This could be the outcome of a scramble competition between gregarious larvae to share resources (Godfray, 1994). The possibility of aggressive behaviour by the first-instar larvae of a gregarious species could explain why egg clutches were larger than the number of offspring in mass rearing of *M. dorsiplana*. In fact, when the parasitized hosts are superparasitized, aggressive encounters between the pteromalid first-instar larvae of *M. dorsiplana* are likely due to their great mobility and well-developed mandibles. In the following phase, although the larvae are immobile and unarmed (personal observations), it is also possible that some brood reduction could occur in hosts containing a large number of gregarious larvae due to over-crowding (Pexton and Mayhew, 2001, Pexton et al., 2003).

In a rearing population cage of *M. dorsiplana* with a density of 1 to 5 females per seed, when one host was parasitized per seed, the sex ratio tended towards a balance of sons (4.12 ± 0.39) and daughters (3.84 ± 0.28), in contrast to the ratio observed with a density of one egg-laying female per host (1 son and 7 daughters) (Rojas-Rousse, 2006). This increase of sons has also been observed in previous experiments with two or three egg-laying *M. dorsiplana* females per host, where the distribution of the associations of 1, 2, 3 or X sons with 1, 2, 3 or X daughters indicates that the common patriline is 2 sons and 8 daughters (Rojas-Rousse, 2006; Stevoux, 1997). The same pattern has been observed among the gregarious pteromalid *Dinarmus vagabundus*, a parasitoid of *C. maculatus*: increasing the density of egg-laying females from one to three per host leads to a greater increase of sons than daughters, the sex ratio (♂/♀) increasing from 0.33 to 1 (Rojas-Rousse *et al*., 1983). Different models have shown the influence of parasitoid density on host-parasitoid population dynamics through local mating competition (LMC) (Hamilton, 1967), the number of female offspring per host being influenced by the density of ovipositing females (Hardy and Ode, 2006). The constraints of mass rearing *M. dorsiplana* in a population cage might prevent the precise application of Hamilton's LMC theory. Some of these constraints need to be tested to understand better the observed fluctuations of the sex ratio of *M. dorsiplana*. For example, asymmetrical mate competition between the broods of different females could occur in a mass-rearing population cage, and females might visit and lay eggs sequentially on different hosts, producing different sex ratios in a patch (Shuker and West, 2004; Shuker et al., 2005). The dispersion of *M. dorsiplana* males from their natal patch before mating has frequently been observed due to the gregarious nature of the hosts in a patch (Jervis and Copland, 1996; Gu and Dorn, 2003), which raises the likelihood of a partial local mating competition in this species.

Studies of the nutritional balance during the development of the gregarious ectoparasitoid *D. vagabundus* have shown that the mean weight of both sexes decreases significantly at higher larval densities (Rojas-Rousse et al., 1988). In a population rearing cage with a high level of ovipositing *M. dorsiplana* females per host, the mean weights of adults emerging from a parasitized host were negatively correlated with egg-clutch size, the larger the egg clutch, the lower the weight. As in other parasitoid species, the different egg-clutch sizes laid by *M. dorsiplana* females might have a considerable impact on offspring fitness (Bezemer et al., 2005; Elzinga et al., 2005;  Milonas, 2005;  Traynor and Mayhew, 2005 a and b).

Overall, this biological information about the newly discovered pteromalid *Monoska dorsiplana* in Latin America indicates that this native gregarious parasitoid could be a promising resource for the biological control of bruchid beetles. When climatic conditions become favourable, the *C. maculatus* bruchid population in storage structures increases rapidly over successive generations (Ouedraogo et al., 1996). To determine whether *M.dorsiplana* could be used as a natural enemy to control this increase in storage systems, its action during regular intervals of introduction need to be analysed after ascertaining that it can move around inside experimental storage systems and locate its hosts, even when these are scarce.

**Acknowledgments**

This article has been read and corrected by Inter-Connect, translation and proof-reading services.

*Table 1.* Distribution of parasitized and non-parasitized hosts in a global set of 200 seeds (40 seeds per day for 5 days). Each seed was opened to observe whether the host was parasitized or not.

| | **Total seeds** | **Presented hosts** | **Parasitized hosts** | **Non-parasitized hosts** |
|---|---|---|---|---|
| **1 host per seed** | 91 | 91 | N=91 91/91= **1** | 0 |
| **2 hosts per seed** | 95 | 190 | N=113 113/190= **0.59** | 77 |
| **3 hosts per seed** | 14 | 42 | N=12 12/42 = **0.28** | 30 |
| **Total** | **200** | **323** | **216** | **107** |

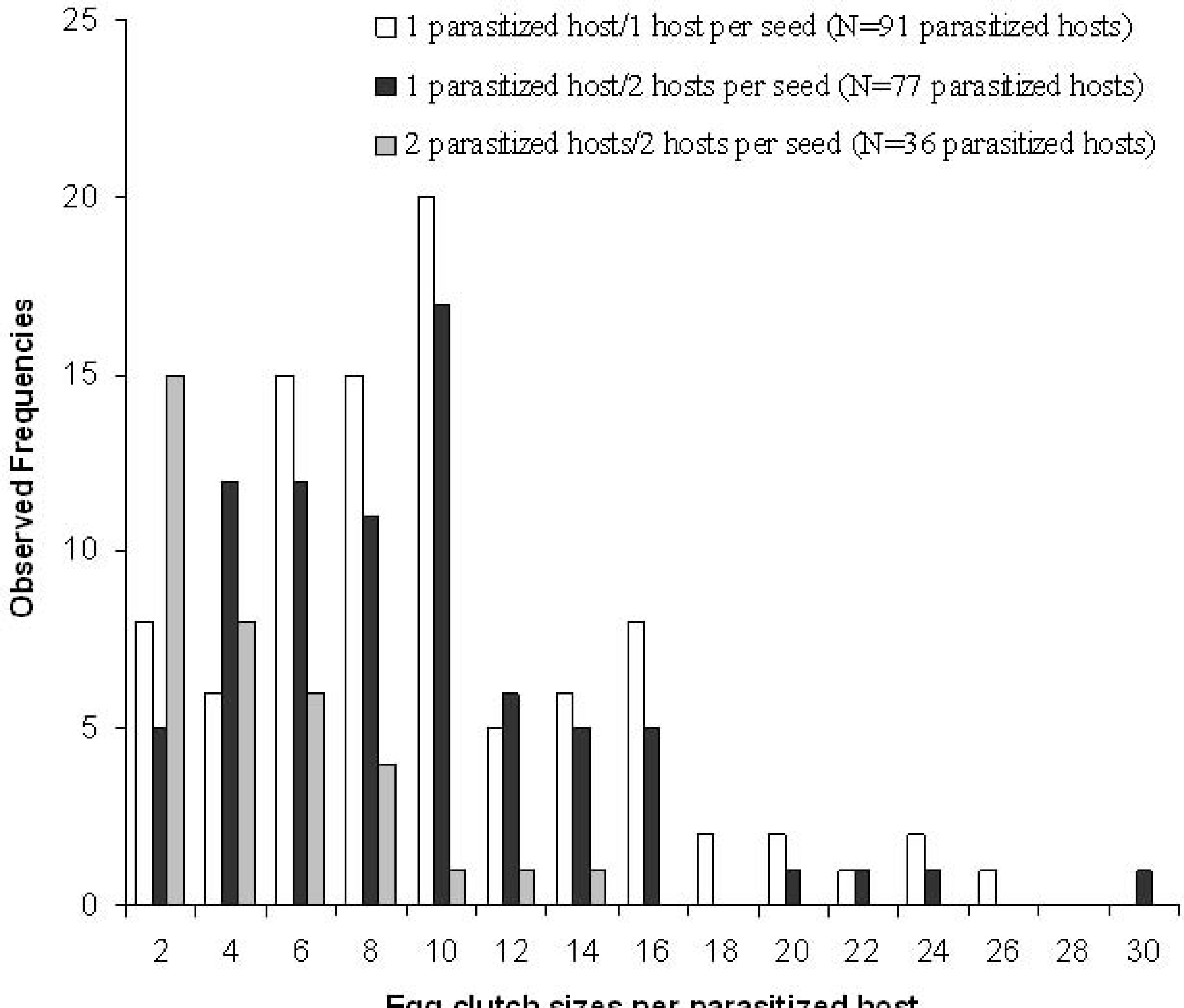

1 parasitized host/1 host per seed (N=91 parasitized hosts)
1 parasitized host/2 hosts per seed (N=77 parasitized hosts)
2 parasitized hosts/2 hosts per seed (N=36 parasitized hosts)
Observed Frequencies
0
5
10
15
20
25
2
4
6
8
10
12
14
16
18
20
22
24
26
28
30

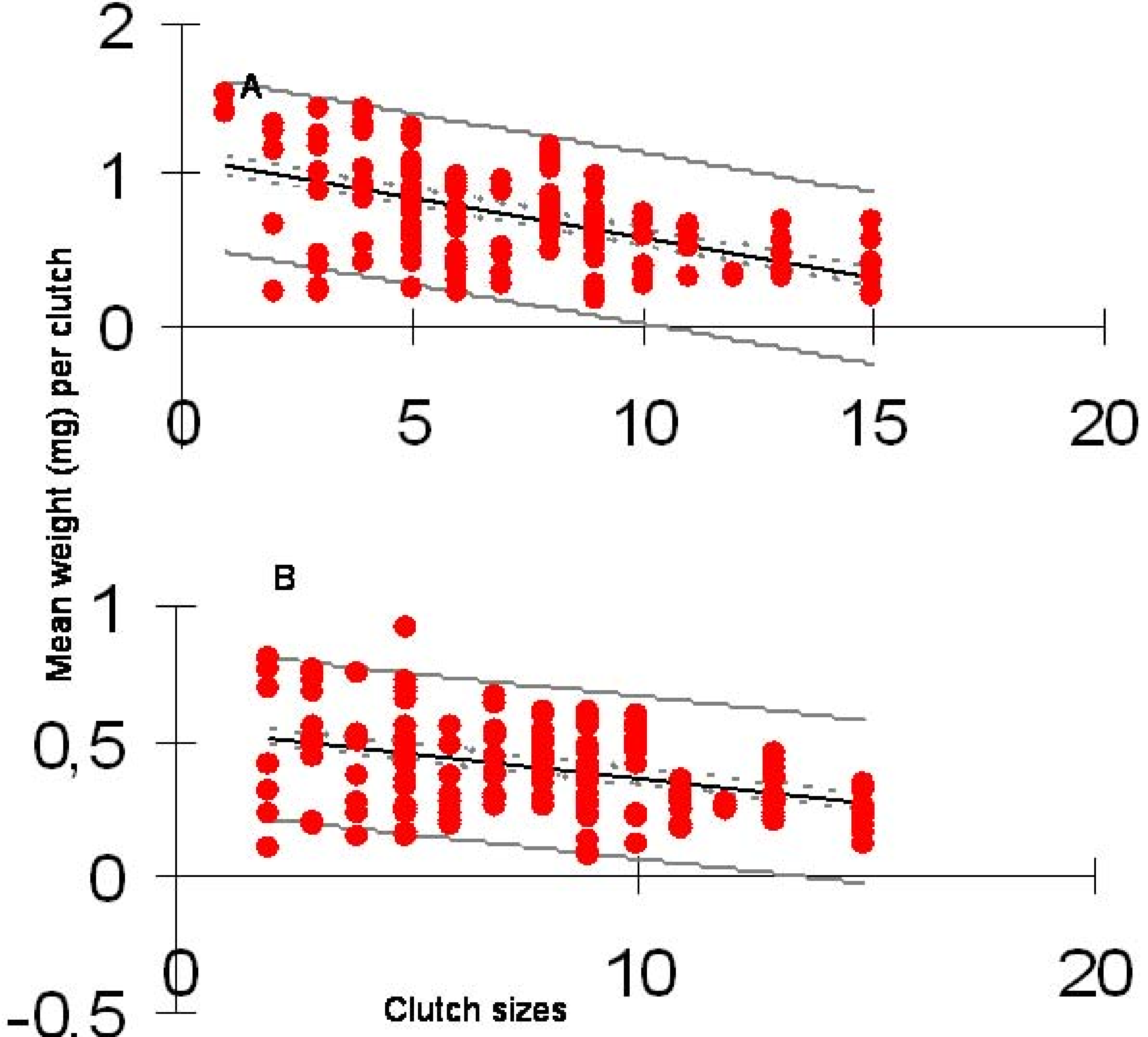